\documentstyle[preprint,aps,prb]{revtex}
\begin{document}
\draft
\title{Magnetic Properties of Epitaxial and Polycrystalline
Fe/Si Multilayers}
\author{A. Chaiken and R.P. Michel}
\address{Materials Science and Technology Division\\Lawrence
Livermore National Lab\\Livermore, CA 94551}
\author{C.T. Wang}
\address{Department of Materials Science and
Engineering\\Stanford University\\Palo Alto, CA 94305}

\maketitle

\begin{abstract}

Fe/Si multilayers with antiferromagnetic interlayer coupling have been
grown via ion-beam sputtering on both glass and single-crystal
substrates.  High-angle x-ray diffraction measurements show that both
sets of films have narrow Fe peaks, implying a large crystallite size
and crystalline iron silicide spacer layers. Low-angle x-ray
diffraction measurements show that films grown on glass have rougher
interfaces than those grown on single-crystal substrates.  The
multilayers grown on glass have a larger remanent magnetization than
the multilayers grown on single-crystal substrates.  The observation
of magnetocrystalline anisotropy in hysteresis loops and $(hkl)$ peaks
in x-ray diffraction demonstrates that the films grown on MgO and Ge
are epitaxial.  The smaller remanent magnetization in Fe/Si
multilayers with better layering suggests that the remanence is not
an intrinsic property.

\end{abstract}
\clearpage

The well-established picture of antiferromagnetic coupling in
metal/metal multilayers would have to be extensively modified for
coupling across insulating or semiconducting spacer layers, where the
spacer does not possess a Fermi surface.\cite{bruno,singh} In
particular the temperature dependence of the exchange coupling might
be significantly different in ferromagnet/semiconductor multilayers
where the exchange is mediated by thermally activated
carriers.\cite{bruno}

Unusual temperature-dependent magnetic properties have been reported
for Fe/Si multilayers.  For example, a large increase in the remanent
magnetization has been observed at low-temperature.\cite{mattson} If
the interlayer antiferromagnetic (AF) coupling increases with
decreasing temperature, as in multilayers with metal spacers, one
might expect the remanence instead to {\em decrease} at low
temperature.  Proper interpretation of the remanent magnetization in
Fe/Si multilayers may therefore be important to understanding the
origin of the interlayer coupling in this system.  One way to explore
the origin of the remanence is to compare films of different
crystalline quality.

\section{Experimental Methods}

The films used in this study were grown using ion-beam sputtering
(IBS) in a chamber with a base pressure of 2$\times$10$^{-8}$ torr.
The deposition system is described in more detail
elsewhere.\cite{fesiprb} All samples used in this study were grown at
a substrate temperature of 200$^{\circ}$.  All comparisons between
films grown on glass and single-crystal substrates will be made on
samples which were deposited {\em simultaneously} so as to eliminate
any reproducibility issues.

The substrates used in this study were glass coverslips, MgO (001),
Ge(001) and Al$_2$O$_3$(11\={2}0).  The MgO and Al$_2$O$_3$ substrates
were cleaned according to a recipe reported by Farrow and
coworkers.\cite{farrow} The glass and Ge substrates were rinsed in
solvents.  All films are capped with a 200\AA\ Ge oxidation barrier.
The magnetic and structural properties of the films are stable for at
least one year.

\section{Structural Characterization}

Figure~\ref{epihigh} shows high-angle x-ray diffraction spectra for a
purely (001)-oriented (Fe40\AA/Si14\AA)x60 multilayer grown on
MgO(001) and a purely (011)-oriented multilayer grown simultaneously
on glass.  While multiple superlattice satellites are observed around
the Fe(002) peak for the film grown on MgO, the film on glass has only
one peak corresponding to (011)-textured growth.  A single
superlattice satellite is typically observed in multilayers on glass
on the low-angle side in keeping with previous
observations.\cite{foiles2} An estimate of the crystallite sizes in
these films can be derived using the Scherrer formula.  This analysis
gives a coherence length of 165\AA\ for the film on glass and 188\AA\
for the film on MgO.

Since the crystalline coherence lengths of these films are similar,
the presence of high-angle satellites in the film on MgO must be due
to better layering.  The small-angle x-ray scattering data shown in
Figure~\ref{epilow} confirm this hypothesis.  The multilayer on MgO
has 4 low-angle peaks, indicating a moderate degree of composition
modulation.  The multilayer on glass shows only two relatively broad
peaks, indicating larger interfacial roughness and less order in the
layering.  The low-angle x-ray spectra are consistent with rocking
curves which are only about 1$^{\circ}$ wide for films grown on MgO
and $\rm Al_2O_3$ but are typically 10$^{\circ}$ to 15$^{\circ}$ wide
for films on glass.  The bilayer periods determined from the low-angle
peak positions are (41.0 $\pm$ 0.1)\AA\ for the MgO film and (40.9
$\pm$ 0.2)\AA\ for the glass film, the same within experimental error.

$\phi$ scans of the MgO and Fe [110] peaks for the film on the MgO
substrate (not shown) demonstrate that it is oriented in-plane.  While
$\theta$-2$\theta$ scans for (011)-oriented multilayers grown on $\rm
Al_2O_3$ substrates (not shown) also show multiple high-angle
superlattice satellites, the $\phi$ scans for this film indicate only
weak orientation in-plane.

The shape of the high-angle peaks and their superlattice satellites
are described by a well-known theory.\cite{fullerton3} Application of
this theory to the Fe/Si multilayers is difficult because the iron
silicide lattice constant, the thickness of the remaining pure Fe and
the thickness of the iron silicide spacer can be estimated only
roughly.  A precise determination of the silicide lattice constant
should make a quantitative analysis of these satellite features
possible.

\section{Magnetic Characterization}

Figure~\ref{bvsh}a shows magnetization curves for 60-repeat
(Fe40\AA/Si14\AA) multilayers grown simultaneously on glass and
MgO(001).  The saturation fields H$\rm _s$ appear to be similar for
the two films.  On the other hand, the remanent magnetization is 58\%
for the film on glass and 7\% for the film grown on MgO.  A remanence
as low as 1\% has been observed for other multilayers grown on MgO
substrates.  SQUID magnetometer data taken up to higher fields gives a
saturation field of 9.75 kOe for the multilayer on MgO at room
temperature.  Assuming for a moment that the interlayer coupling is
purely bilinear in nature, a well-known formula relates the saturation
field to the AF coupling strength: A$_{12} \, =
\, H_s M_s t_{Fe} / 4$ where M$\rm _s$ is the saturation magnetization
and t$\rm _{Fe}$ is the thickness of an individual Fe
layer.\cite{krebs} Use of this equation with H$\rm _s$ = 9.75 kOe and
the measured magnetization M$_s$ = 1271 emu/cm$^3$ gives A$\rm _{12}$
= 1.2 erg/cm$^2$.  This AF coupling value is comparable in size to the
coupling measured in metal/metal multilayers multilayers\cite{parkin}.

Figure~\ref{bvsh}b shows magnetization curves for
Fe100\AA/Si14\AA/Fe100\AA trilayer films grown on Al$_2$O$_3$
(11\={2}0), Ge(001), and MgO(001) substrates.  All three of the
magnetization curves in Fig.~\ref{bvsh}b were taken with the field
applied along an Fe(100) easy direction.  Significant in-plane
anisotropy of the magnetization curves occurs for the films on the Ge
and MgO substrates, similar to what has been observed for Fe/Cr/Fe
trilayers.\cite{krebs} The observation of magnetocrystalline anisotropy
in the film on MgO but not in the film grown on Al$_2$O$_3$ is
consistent with expectations from the $\phi$ scans, which show that
the in-plane orientation of the film on MgO is much stronger.

Figure~\ref{bvsh}b once again demonstrates that the degree of
remanence in Fe/Si multilayers is strongly related to the quality of
layering.  While the remanent magnetization of the epitaxial trilayers
on Ge and MgO is only about 5\% of the saturated value, the remanence
of the polycrystalline trilayer on Al$_2$O$_3$ is close to 50\%.  The
remanent magnetization of the trilayers on Ge and MgO is about 5\% in
the in-plane hard direction (H $\|$ Fe(110)) as well.

A SQUID magnetometer has been used to measure the magnetization curves
of the IBS-grown Fe/Si multilayers at lower temperatures.\cite{michel}  The
temperature dependence of the remanent magnetization of these films is
similar to that reported by other authors.\cite{mattson}

\section{Discussion}

At the moment it is not possible to tell why the in-plane
ordering of the films grown on Al$_2$O$_3$(11\={2}0) is inferior to
that grown on the (001) MgO and Ge substrates.  The difficulty with
the Al$_2$O$_3$ growth may have to do with the 6$^{\circ}$ miscut of
the substrates, or it may be due to an intrinsic difficulty with (011)
growth of the Fe/Si multilayers.  Previous work has shown that AF
coupling in Fe/Si multilayers is dependent upon formation of a
metastable iron silicide spacer layer phase.\cite{fesiprb} The
possibility exists that the spacer silicide does not grow well on Fe
in the (011) orientation.  This question can be answered only by
further growth studies on better (011) substrates and careful
structural characterizations.

A related question is whether the larger remanent magnet moment in the
(011)-textured films might be due to a fundamental difference in
magnetic properties from the (001)-textured films.  Because a
46-repeat Fe/Si multilayer grown on Al$_2$O$_3$ has a remanence of
only about 10\%, this is unlikely.  Undoubtedly the trilayer on
Al$_2$O$_3$ has a higher remanence than the multilayer because the
thinner film is more greatly impacted by the poor substrate surface
quality.  The staircase morphology caused by the 6$^{\circ}$ miscut of
this Al$_2$O$_3$ substrate may lead to wavy interfaces between the Fe
and iron silicide films or to pinholes through the silicide layers.
Wavy interfaces can cause increased magnetostatic coupling or even
biquadratic coupling,\cite{slonczewski} both of which would tend to
increase the remanence.  The large remanence of the multilayers grown
on glass substrates is likely also due to pinholes or magnetostatic
coupling.

Pinhole-induced may explain the unusual temperature dependence of the
remanence.  (Magnetostatic coupling is expected to be approximately
temperature-independent.)  Fe atoms in bridges through the silicide
spacer layers are expected have a reduced Curie temperature.  A larger
remanence at low temperature therefore makes sense if the remanence is
derived from pinhole coupling and is not an intrinsic effect.  Low
Curie-temperature material may also be present in the iron silicide
spacer layer or in at the iron/iron silicide interfaces.

By growing on a number of substrate materials and by using different
deposition conditions, Fe/Si multilayers have been prepared with a
varying degree of ordering.  A large amount of accumulated evidence
demonstrates that high remanence of the magnetization curves in Fe/Si
multilayers is associated with interface roughness.  The remanence is
therefore not likely to be related to unusual exchange coupling but
instead to originate from defects, perhaps pinholes through the
silicide spacer layer.  Since the remanent magnetization is caused by
extrinsic effects, future studies should concentrate instead on
measurements of the saturation field of the magnetization curves in
order to learn more about the interlayer coupling.

\bigskip

\noindent We would like to thank E.E. Fullerton, J.A. Borchers,
R.M. Osgood III and Y. Huai for helpful discussions, and B.H. O'Dell and
S. Torres for technical assistance.  Part of this work was performed
under the auspices of the U.S. Department of Energy by LLNL under
contract No. W-7405-ENG-48.

\clearpage

\begin{figure}
\caption{High-angle x-ray diffraction spectra from two
(Fe40\AA/Si14\AA)x60 multilayers grown on different substrates.
High-angle x-ray peaks give information about lattice constants and
stacking within the layers.  The multilayer grown on MgO(001) has an
Fe(002) peak with 4 satellites.  The multilayer simultaneously grown
on glass has only an Fe(011) peak.}
\label{epihigh}
\end{figure}

\begin{figure}
\caption{Low-angle x-ray diffraction spectra for the same two
(Fe40\AA/Si14\AA)x60 multilayers.  Low-angle x-ray peaks give
information about the bilayer period and layering of the multilayer.
The multilayer grown on MgO has four relatively narrow peaks.  The
multilayer grown on glass has only two broad peaks, indicating a
greater degree of interface roughness.}
\label{epilow}
\end{figure}

\begin{figure}
\caption{Magnetization curves for Fe/Si multilayers grown on various
substrates.  The magnetization data is normalized to the highest
measured value in order to facilitate comparison of the shape of the
two data sets.  a) Hysteresis loops for the (Fe40\AA/Si14\AA)x60
multilayers simultaneously grown on MgO (001) and glass.  The
remanence is lower for the film grown on MgO. b) Hysteresis loops for
(Fe100\AA/Si14\AA)x2 multilayers (trilayers) grown on Al$_2$O$_3$
(11\={2}0), Ge(001), and MgO(001).  All data are taken with the
applied magnetic field along the Fe (100) easy direction.}
\label{bvsh}
\end{figure}

\end{document}